\documentclass[floatfix,aps,pre,amsmath,amssymb,amsfonts,superscriptaddress,reprint]{revtex4-1}
\usepackage{bm}
\usepackage{graphicx}
\usepackage{hyperref}
\usepackage{xcolor}
\usepackage{soul}

\begin{document}

\title{Self-trapping of active particles with non-reciprocal interactions in disordered media}

\author{Rodrigo Saavedra}
\email{rsaavedra@cicese.mx}
\affiliation{%
    Centro de Investigaci\'on Científica y de Educaci\'on Superior de Ensenada, Carretera Ensenada-Tijuana 3918, 22860, Ensenada, M\'exico %
}
\author{Fernando Peruani}
\email{fernando.peruani@cyu.fr}
\affiliation{%
    Laboratoire de Physique The\'eorique et Mod\'elisation, UMR 8089, CY Cergy Paris Universit\'e, Cergy-Pontoise, France %
}

\date{\today}

\begin{abstract}
    We study systems of active particles, whose perception is constrained by a vision cone, that  are attracted to other particles and repelled from static obstacles. %
    We report a novel self-trapping mechanism: active particles with non-reciprocal attraction form particle chains, which eventually become closed loops that shrink around one or many obstacles. %
    These closed loops act as effective aggregation centers. %
    Long-lived, self-organized closed loops require to enclose obstacles to exist. %
    Furthermore, we find that closed loops that initially exhibit local polar order, transition to a nematic state as they absorb more particles. %
    The unveiled mechanism corresponds to a pinning behavior that strongly hinders  particle diffusion. %
    In short, closed loops dominate the large-scale properties of active systems with non-reciprocal attraction in disordered media. %
\end{abstract}

\maketitle

\section{Introduction}
Most examples of natural systems, if not all, where collective motion occurs in the wild, take place in heterogeneous media. %
Examples can be found at all scales. %
Microtubules driven by molecular motors form complex patterns inside the cell where the space is filled by organelles and vesicles~\cite{bruno2011mechanical,barlan2017microtubulebased}. %
Bacterial colonies exhibit complex collective behaviors, e.g. swarming in heterogeneous environments such as the soil~\cite{partridge2013swarming,yang2017influence} or highly complex tissues such as in the gastrointestinal tract~\cite{altshuler2013flowcontrolled,figueroa-morales2015living,nguyen2020environmental}. %
At a larger scale, herds of mammals migrate long distances traversing rivers, forests, etc.~\cite{seidler2015identifying,leclerc2021determinants}. %
In (vectorial) active matter, the influence of static obstacles, or quenched disorder, on the emergent collective behavior, has been the focus of many recent studies~\cite{rahmani2021topological, codina2022small,chepizhko2013diffusion,chepizhko2013optimal, peruani2018cold, mokhtari2019dynamics,zhang2021polymer,theeyancheri2022migration}. %
For example, when velocity alignment is considered, it has been found that a single small obstacle can have a dramatic impact in the large-scale dynamics of polar flocks, leading to flow reversals and chaotic dynamics~\cite{codina2022small}. %
In the presence of several randomly placed obstacles, the system becomes disordered and particles aggregate into independent clusters~\cite{chepizhko2013diffusion,chepizhko2013optimal}. %
Moreover, it was shown that the long-range order (LRO) that emerges in two-dimensional, spatially homogeneous systems~\cite{vicsek1995novel}, becomes quasi-long-range (QLRO)~\cite{chepizhko2013optimal}, with particle motion  constrained to spontaneously formed streams that go through the disorder media~\cite{peruani2018cold}. %
Finally, for active polymers, it was found that porous media either rectify or segregate the  polymers~\cite{mokhtari2019dynamics,zhang2021polymer,theeyancheri2022migration}. %

On the other hand, non-reciprocal interactions pertains to a new class of active matter that can be found in system interacting via chemotaxis, hydrodynamic forces, or quorum-sensing~\cite{dinelli2023nonreciprocity}. %
The impact of non-reciprocal interactions in active systems has been recently explored in theoretical studies~\cite{barberis2016largescale,peruani2017hydrodynamic, durve2018active,you2020nonreciprocity,fruchart2021nonreciprocal,kreienkamp2022clustering,knezevic2022collective}. %
Particularly relevant, in terms of applications and conceptually, are those model where agents navigate using visual information that is restricted by a vision cone~\cite{barberis2016largescale,newman2008effect,stengele2022group,qi2022emergence,negi2022emergent,saavedra2024swirling}. %
One fundamental lesson learned from these models is that flocking behavior can emerge in the absence of velocity alignment -- present for instance in Vicsek model -- and as result of non-reciprocal attraction~\cite{barberis2016largescale}. %
This concept was  successfully applied to describe the collective dynamics of sheep~\cite{ginelli2015intermittent, gomez2022intermittent} and to induce cohesive group formation in swarms of light-activated colloids~\cite{lavergne2019group,qi2020group,chen2022collective}. %

In this work, we study, for the first time, the impact an heterogeneous medium, i.e. quenched disorder~\cite{chepizhko2013diffusion,chepizhko2013optimal, sandor2017dynamic, martinez2018collective, toner2018swarming,kumar2022active,duan2021breakdown,jacucci2024patchy} has on the collective dynamics of active particles with non-reciprocal attractive interactions. %
Specifically, we investigate a new mechanism of self-trapping of active particles that consists of closed loops around static obstacles. %
This behavior is found to emerge in system of active particles with a visual-like perception that are attracted to conspecifics and repelled from static obstacles. %
The emergent closed loops consist of active particles whose  velocity is (locally) tangential to the loop. %
Once formed, closed loops can absorb incoming particles, but particles in the loop can also escape by (angular) diffusion. %
Our analysis shows that self-trapping -- occurring for  intermediate values of vision cone angle -- corresponds to a pinning behavior that strongly hinders particle diffusion. %
Active particles  coalesce into the spontaneously formed closed loops. %
Our study sheds light on the dynamics of active matter with non-reciprocal interactions in complex environments. %

\section{Model}
\begin{figure}[b!]
    \includegraphics[width=\linewidth]{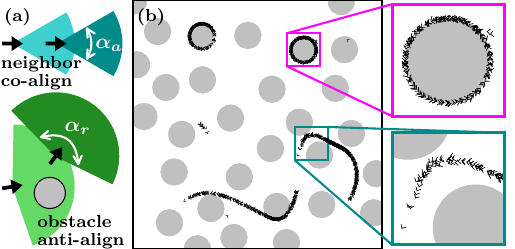}%
    \caption{\label{fig:1}%
        (a)~Illustration of alignment interactions given by~Eqs.~\eqref{eq:alignment_terms}. %
        Each active particle perceives neighbors within a narrow vision cone of angle~$\alpha_a$, and obstacles within a  vision cone of angle~$\alpha_r$. %
        (b)~Representative snapshot for a configuration of $N=900$ particles and $N_o=30$ static obstacles. %
        Simulation parameters are~$\alpha_a=0.3\pi$, $\alpha_r=0.6\pi$, $\sqrt{2D_{\theta}}=0.3$. %
    }
\end{figure}

We consider a system of $N$ motile particles and $N_o$ static obstacles distributed inside of a square box of length~$L$ with periodic boundary conditions. %
The motion of the $i$-th active particle is governed by: %
\begin{subequations}
    \begin{align}
        \dot{\bm{r}}_i &= v_0\bm{e}_i \label{subeq:position_eom},\\
        \dot{\theta}_i &= \tau_i^\mathrm{att} + \tau_i^\mathrm{rep} + \xi_i. \label{subeq:theta_eom}
    \end{align}
    \label{eq:eom}
\end{subequations}
Here $v_0$ and $\bm{e}_i=(\cos\theta,\sin\theta)^T$ are the self-propulsion magnitude and direction, respectively, $\xi_i$ is a delta-correlated noise, and $D_\theta$ is the rotational diffusion coefficient. The terms $\tau_i^\mathrm{att}$, $\tau_i^\mathrm{rep}$ correspond to  torques induced by neighboring particles $j$ and obstacles $o$, respectively. They are given by %
\begin{subequations}
    \begin{align}
        \tau_i^\mathrm{att} &=  \frac{\gamma^\mathrm{att}}{n^\mathrm{att}}\sum_{j\in{V}_i^\mathrm{att}}\sin(\beta_{ij}-\theta_i), \\
        \tau_i^\mathrm{rep} &= -\frac{\gamma^\mathrm{rep}}{n^\mathrm{rep}}\sum_{o\in{V}_i^\mathrm{rep}}\sin(\beta_{io}-\theta_i),
    \end{align}
    \label{eq:alignment_terms}
\end{subequations}
with $\beta_{ij}$ the polar angle of the vector $\bm{r}_{ij}=\bm{r}_j-\bm{r}_i$, and similarly for $\beta_{io}$. %
Coefficients $\gamma^\mathrm{att}$, $\gamma^\mathrm{rep}$ are attraction and repulsion torque strengths, respectively. %
Importantly, $V_i^\mathrm{att}$ and  $V_i^\mathrm{rep}$ are the vision cones related to interactions with other active particles and obstacles, respectively. %
The vision cone~$V_i^\mathrm{att}$ is characterized by an angle~$\alpha_{a}$, while $V_i^\mathrm{rep}$ by an angle $\alpha_{r}$. %
Furthermore, both vision cones, $V_i^\mathrm{att}$ and $V_i^\mathrm{rep}$, share the same perception horizon $r_c$;  see~Fig.~\ref{fig:1}a. %
The terms $n^\mathrm{att}$ and $n^\mathrm{rep}$ correspond to the number of neighbors within each vision cone. %
Note that the phenomena described below, including self-trapping, is reported mainly for $\alpha_a=\alpha_r=\alpha$. %
However, we performed also  analysis  relaxing this constraints and assuming $\alpha_a \neq\alpha_r$. %
We consider a configuration of randomly distributed obstacles that leads to an average distance $d$ between obstacles. %
As we increase the number of obstacles $N_o$, $d$ decreases. %
We perform particle-based numerical simulations by solving Eq.~\ref{eq:eom} for~$N=900$ particles that initially are  randomly distributed within a square box of side length $L=30$. %
We fix self-propulsion speed~$v_0=1$, torque strengths~$\gamma^\mathrm{att}=\gamma^\mathrm{rep}=\gamma$ with~$\gamma=5$, and a vision cone horizon~$r_c=1.5$. %
The P\'eclet number is defined as~$\mathrm{P_e}=v_0/(\sigma{D}_\theta)$, with $\sigma=2r_c$. %
    We choose two different sets of parameters for~$(\alpha,D_\theta)$. %
    One set with $\alpha=0.3\pi$ and $\sqrt{2D_{\theta}}=0.3$ ($\mathrm{P_e}\approx{15}$) where particles are known to aggregate into \emph{polar filaments}. %
Another set with $\alpha=0.6\pi$ and $\sqrt{2D_{\theta}}=0.8$ ($\mathrm{P_e}\approx{2}$) where particles form percolating \emph{nematic bands}, see Ref.~\hbox{\cite{barberis2016largescale}}. %

\section{Results}
\subsection{Polar filaments}
In a wide range of parameters, starting from a random distribution of positions and velocities, active particles self-organize into polar filaments; Fig.~\ref{fig:1}b. %
A polar filament consists of a chain of particles moving altogether in the same direction. %
The direction of motion of the chain is determined by the motion of the particle at the head of the structure, which becomes an ``incidental leader". %
The leader is then followed by a tail of particles. %
A particle~$j$ in the tail reorients its direction of motion~$\bm{e}_j$ towards the position of the active particles located within its (particle interaction) vision cone~$V^\mathrm{att}_j$. %
The leader, on the other hand, does not perceive any particle in its vision cone, and thus its  orientation~$\bm{e}_i$ is affected exclusively by rotational diffusion~$D_{\theta}$. %
Particles are scattered away from  obstacles that they perceive within their (obstacle interaction) vision cone~$V^\mathrm{rep}_i$. %
Particles can escape a polar filament due to a strong fluctuation of their self-propulsion orientation, due to the encounter with an obstacle, or due to the encounter with other particle. %
Collisions of several filaments can lead to aggregation into a single  structure. %

\begin{figure}[t!]
    \includegraphics[width=\linewidth]{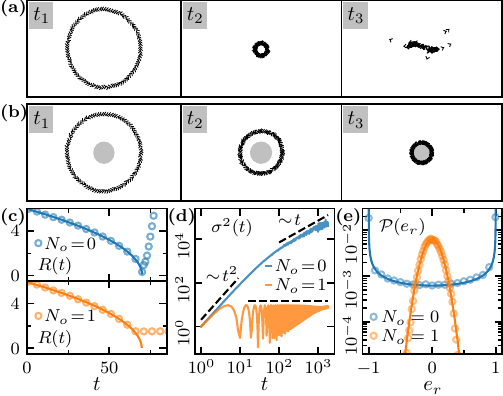}%
    \caption{\label{fig:2}%
        (a)~Without an obstacle, the loop shrinks down to a size~$r<r_o$, then fragments into several polar filaments that propagate radially out of the center. %
        (b)~With an obstacle the loop shrinks down to a size $r=r_o$ and stabilizes around the obstacle. %
        (c,d)~Time evolution of dynamic parameters corresponding to configurations in~(a,b), namely loop radius~$R(t)$, and mean squared displacement~$\sigma^2(t)$. %
        Solid lines in~(c) correspond to Eq.~\eqref{eq:loop_radius}. %
        (e)~Probability density of finding a particle with radial orientation~$e_r$ with respect to the center of the configuration at initial time. %
        Solid lines correspond to $\mathcal{P}(e_r)\sim\exp(-e_r^2)$ for $N_o=0$, and to $P(e_r)\sim{1}/\sqrt{1-e_r^2}$ for $N_o=1$. %
    }
\end{figure}

A polar filament can eventually form a (polar) closed loop,  when the incidental leader of a motile filament starts following its own tail. %
Loop formation is strongly enhanced by the presence of obstacles that promote the fast reorientation of the leader. %
From simulations we observe that filaments navigate while trying to avoid obstacles, and furthermore they show to self-trap around some of them, see~Fig.~\ref{fig:1}b. %
Below, we show that initially formed polar loops become over time nematic loops, with some particles rotating clockwise, and some other particles rotating counter-clockwise. %
This results from fluctuations in the moving direction, but also due to the arrival of other active particles that join the loop. %
Importantly, once formed, the loop can shrink  down or grow. %
The closed loop behaves as a elastic band, whose physics and dynamics should not be confused with effective line tension that emerges on interfaces in active Brownian particles systems~\hbox{\cite{bialke2015negative}.} %
If a loop is formed and shrinks without containing any obstacle, its radius $R$ reaches $v_0/\gamma^\mathrm{att}$ -- that is the smallest the equation of motion allows -- and the loop breaks into several polar filaments propagating radially outwards of the loop center, see~Fig.~\ref{fig:2}a. %
However, if one or several obstacles are located in the inner area of the loop, the loop shrinks until touching the obstacles as a rubber band; see Fig.~\ref{fig:2}b. %

\begin{figure}[t!]
    \centering
    \includegraphics[width=\linewidth]{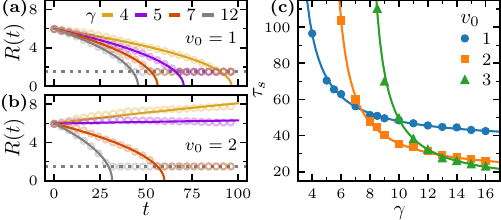}
    \caption{\label{fig:4}%
        Loop radius for several values of the alignment strength~$\gamma$ and self-propulsion speed of %
        (a)~$v_0=1$, and %
        (b)~$v_0=2$. %
        Horizontal dotted line indicates the obstacle size. %
        Solid lines correspond to Eq.~\eqref{eq:loop_radius}. %
        (c)~Shrinking time~$\tau_s=R_0^2/(2\kappa)$ obtained by fitting curves like the ones in~(a,b). %
        Particle speeds $v_0=1,2,3$ shown correspond to P\'eclet numbers $\mathrm{P_e}\approx{15,30,45}$, respectively.
    }
\end{figure}

To quantify the loop behavior, we compute the radius of a closed loop with $N=100$ particles:  %
\begin{equation}
    R^2(t) = \frac{1}{N}\sum_{i=1}^{N}(\bm{r}_i(t)-\bm{r}_\mathrm{cm}(t))^2\,,
\end{equation}
where $\bm{r}_\mathrm{cm}$ corresponds to the center of the loop. %
Results in~Fig.~\ref{fig:2}c show that the radius~$R$ of a closed loop initially decreases in the presence or absence of obstacles. %
{It is possible to derive an equation for the temporal dynamics of $R$ (that can result in either the shrinking or growing of~$R$). %
    Consider a ring of particles of radius $R$ at time $t$ and let us focus on a single particle located at the ``north pole'' of the circle. %
    Note that we expect the behavior of all particles in the circle to be identical. %
    This particle has non-zero velocity components, i.e. $\dot{x}$ and $\dot{y}$, which are given by Eq.~\hbox{\eqref{subeq:position_eom}}. %
    It is only $\dot{y}$ the component that affects instantaneously (at order $dt$) the distance of the particle to the center of the circle, and thus $\dot{y}=\dot{R}$. %
    Assuming $|\theta|\ll 1$, and using Eq.~\hbox{\eqref{subeq:position_eom}}, then $\dot{R}=\dot{y}= - v_0 \theta$. %
    On the other hand, according to Eq.~\hbox{\eqref{subeq:theta_eom}}, $\dot{\theta} = \gamma/n \, \sum_j \sin(\beta_j - \theta)$, ignoring the noise term and assuming that the particle does not perceive the obstacle. %
    Let us consider, for simplicity, that the particle perceives only one neighbor, and assume a fast angular dynamics such that $\dot{\theta}=0$ and thus $\theta = \beta$. %
From the definition of radius of curvature, $R=|\Delta{\ell}/ \Delta \phi|$, where $\Delta{\ell}$ is the variation of arc length and  $\Delta \phi$ the variation of tangential angle, it is evident that $\beta \sim \Delta \phi$ and thus, $\beta = c/R$, with $c$ is a constant.} %
Putting all together, we obtain: %
\begin{equation}
    \dot{R} = -\frac{\kappa}{R} \, ,
\end{equation}
where $\kappa$  is a constant that depends on $v_0$ and $\gamma$. %
The solution of this equation is: %
\begin{equation}
    R(t) = R_{0}\sqrt{1-t/\tau_s} \, ,
    \label{eq:loop_radius}
\end{equation}
as shown in Fig.~\ref{fig:2}(c) and \ref{fig:4}, here $R_0$ is the initial radius, $\tau_s=R_{0}^2/(2\kappa)$ is the time it takes a loop to shrink down to its minimum size (approximately zero). %
Without the obstacle, $R$ decreases to a minimum value of~$R\approx{0}$, then it diverges at longer times when the loop has lost cohesion forming propagating polar filaments. %
When the loop self-traps around the obstacle, its radius reaches a constant value proportional to the size of the obstacle,~$R\approx{r_o}$. %

Moreover, we calculate the mean-squared displacement of the particles: %
\begin{equation}
    \sigma^2(t) = \frac{1}{N}\sum_{i=1}^{N}
    \langle(\bm{r}_i(t+t_0)-\bm{r}_i(t_0))^2\rangle,
\end{equation}
where the average is over the initial time~$t_0$. %
$\sigma^2$ is found to transition from ballistic to diffusive regime in the case without an obstacle, and displays a plateau in the case with an obstacle. %
We conclude that self-trapping corresponds to a pinning mechanism that drastically impacts the transport properties of the system. %

Finally, to quantify the average orientation of the particles in the loop, we compute the radial orientation component~${e_r}_i=\hat{\bm{e}}_i\cdot\hat{\bm{r}}_i = \cos(\Delta_i)$ of each particle with respect to the center of mass; here~$\hat{\bm{r}}_i$ is a unitary radial vector with origin at the center of the configuration at initial time. %
Here,~${e_r}_i=1$ indicates particles are radially outward oriented, and~${e_r}_i=-1$ are radially inward oriented, wheres~${e_r}_i=0$ corresponds to tangentially oriented particles. %
We obtain the probability density~$\mathcal{P}(e_r)$, see~Fig.~\ref{fig:2}e. %
We find that, in the case with an obstacle,~$\mathcal{P}(e_r)$ is centered around~$e_r=0$ (i.e. $\Delta$ is centered around $\pi/2$) and follows a Gaussian functional form $\mathcal{P}(e_r) \sim \exp(-e_r^2)$. %
This shows that trapped particles have an orientation vector mostly tangential to the obstacle with slight deviations corresponding to the width of~$\mathcal{P}$. %
Deviations result from fluctuations due to rotational noise. %
In the case without an obstacle, $\mathcal{P}(e_r) \sim 1/\sqrt{1-e_r^2}$ as expected for a fully random distribution of $\Delta$ angles.

We observe the orientational order of trapped loops transitions from being polar to become nematic over time. %
Consider an initially polar closed loop consisting of a certain number of active particles that can be rotating either clockwise~(CW) or counter-clockwise~(CCW). %
In contrast to the Vicsek model, the positional-based alignment defined in~Eq.~\eqref{eq:alignment_terms} does not align orientations (i.e. velocities) of neighboring particles to be the same. %
Local order in this case can be either polar or nematic. %
\begin{figure}[b!]
    \centering
    \includegraphics[width=\linewidth]{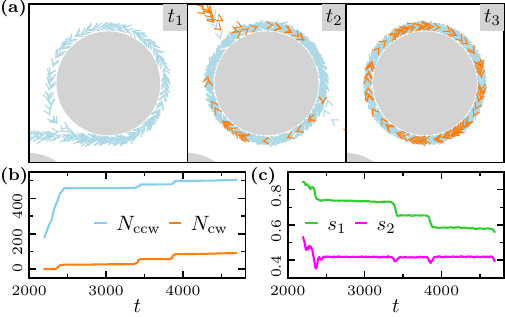}
    \caption{\label{fig:3}%
        (a)~Transition of a loop from initially polar to finally nematic. %
        Time evolution of~%
        (b)~the number of clockwise~$N_\mathrm{CW}$ and counter-clockwise~$N_\mathrm{CCW}$ particles in the loop, and~%
        (c)~the local polar order~$s_1$, as well as local nematic order~$s_2$. %
    }
\end{figure}
In~Fig.~\ref{fig:3}a we show a polar loop formed by~$N_\mathrm{CCW}<N$ particles. %
As the system evolves, more particles can join the loop, and they can join either in the CW or CCW direction. %
Therefore, an initially polar loop can become nematic over time with a certain amount of~$N_\mathrm{CCW}$ and~$N_\mathrm{CW}$ particles. %
In~Fig.~\ref{fig:3}b we show the time evolution of $N_\mathrm{CW}$ and $N_\mathrm{CCW}$ in the loop. %
We observe that the number increases over time as incoming bulky filaments  are absorbed by the loop. %
The loop in this case starts with~$N_\mathrm{CCW}=200$, and~$N_\mathrm{CW}=0$ particles at~$t=2200$. %
At a later time~$t=4700$ the loop consists of~$N_\mathrm{CCW}\approx{500}$ and~$N_\mathrm{CW}\approx{100}$ particles. %
We also calculate the time evolution of the local orientational order~$s_n$ with the formula %
\begin{equation}
    s_n(t) = \bigg|\frac{1}{N}\sum_{i=1}^{N}e^{in\theta_i(t)}\bigg|, %
    \label{eq:polar_nematic_order_parameter}
\end{equation}
computed within a local cutoff radius of~$r_c=1.5$. %
Polar order corresponds to $n=1$, while $n=2$ to nematic order. %
We observe a decrease from~$s_1\approx{0.85}$ to~$s_1\approx{0.6}$ in the final configuration. %
The local nematic order~$s_2$ decreases only slightly from~$s_2\approx{0.5}$ to~$s_2\approx{0.4}$. %

To quantify the shrinking behavior of the loop, we consider an initially nematic loop of $N=100$ particles placed at a distance of~$R_{0}=6$ from the center of a single obstacle. %
For the standard parameters considered, namely~$v_0=1$ and~$\gamma=5$, we obtain that the loop shrinks and gets trapped around the obstacle.  %
We test this scenario for several values of the turning strength~$\gamma$ at~$v_0=1$, and obtain the loop radius~$R$ for each realization. %
Results are shown in~Fig.~\ref{fig:4}a. %
For~$v_0=1$, we observe that~$R$ decreases and saturates at longer times when the loop has stabilized around the obstacle, reaching the minimum size of~$R=r_o$. %
For larger values of~$\gamma$, the loop radius~$R$ shows to decrease more rapidly. %
Furthermore, for~$v_0=2$, we observe that with the small turning speed~$\gamma=4$ shown here, $R$ monotonously increases in time, indicating that the loop does not shrink (but grows instead) in this case, see~Fig.~\ref{fig:4}b. %
For~$\gamma=5$, $R$ remains mostly constant for the simulation time here considered. %
To better understand the loop shrinking, we compute the shrinking time~$\tau_s$ from~$R(\tau_s)\approx{0}$. %
Results in~Fig.~\ref{fig:4}c show that the shrinking time monotonously decreases with~$\gamma$ and it also decreases with $v_0$. %
At larger values of~$\gamma$, $\tau_s$ is not expected to change, as in those cases the torque is already large enough to ensure particles will rapidly turn towards the loop center and the shrinking time will be determined only by the particle speed~$v_0$. %

\begin{figure}[t!]
    \includegraphics[width=\linewidth]{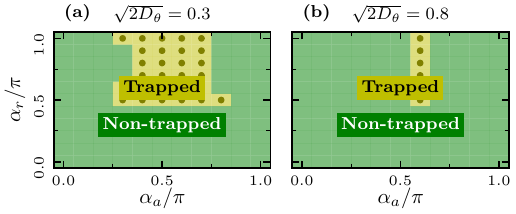}
    \caption{\label{fig:5}%
        Single-loop trapping diagram for different values of~$\alpha_a$ and $\alpha_r$ at %
        (a)~low rotational diffusion ($\mathrm{P_e}\approx{15}$), and %
        (b)~high rotational diffusion ($\mathrm{P_e}\approx{2}$). %
    }
\end{figure}

In the absence of external disturbances, the stability of a particle loop surrounding an obstacle is not guaranteed. %
For example, large values of~$D_\theta$ can also trigger de-trapping, as it directly influences the particle orientation~$\bm{e}$, allowing them to point radially out of the loop configuration. %
When the vision cone angle~$\alpha_a$  is narrow, particles pointing out of the loop configuration do not perceive other neighbors and do not co-align to join the loop anymore, instead they escape the trap. %
This can trigger de-trapping of the whole structure, as the escaping particle can serve as an incidental leader which will be followed by the rest of the particles in the loop. %
To test particle trapping for different parameters of the cone of vision as well as noise strength, we perform simulations of a single loop around one obstacle. %
We take $v_0=1$ and consider several values of~$\alpha_a$ and~$\alpha_r$, both at low and  high values of the noise strength~$\sqrt{2D_{\theta}}$. %
See results in~Fig.~\ref{fig:5}. %
We observe that at low noise, $\sqrt{2D_{\theta}}=0.3$, trapping only occurs for large values of the obstacle perception angle~$\alpha_r\geq{0.5}\pi$, and for intermediate values of the neighbor perception angle~$\alpha_a$. %
For high noise, $\sqrt{2D_{\theta}}=0.8$, we observe trapping only occurs for~$\alpha_a=0.6 \pi$ and large~$\alpha_r>0.5\pi$. %

\begin{figure}[t!]
    \includegraphics[width=\linewidth]{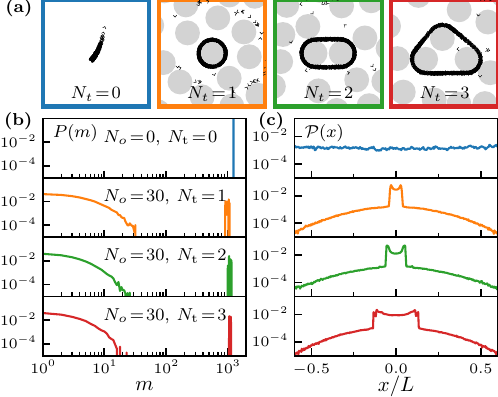}
    \caption{\label{fig:6}%
        (a)~Polar filament found in the case without obstacles, and closed loops trapped around $N_t=1,2,3$ obstacles. %
        (b)~Cluster size distribution~$P(m)$, and %
        (c)~marginal displacement probability density~$\mathcal{P}(x)$ for~$N_o=0,30$. %
        Cases considered correspond to those in panel~(a). %
        For~$P(m)$ a clustering analysis is performed with a cutoff radius of~$r_c=r_o$. %
        For~$\mathcal{P}(x)$, displacements are calculated during a time interval of~$\Delta{\tau}=15$. %
    }
\end{figure}

Loops can be formed around more than a single obstacle and therefore they can be extended in space and contain a large fraction of particles of the system. %
To characterize such behavior, first we obtain the cluster size distribution %
\begin{equation}
    P(m)=\frac{m\,n_m}{N}\,,
\end{equation}
where $m$ is the cluster size, and $n_m$ the number of clusters of size~$m$. %
We perform a clustering analysis with a cutoff radius of~$r_c=1.5$, neighbors are considered to be part of a single cluster only when they are close together a distance~$r<r_c$. %
The clustering analysis is averaged over a time interval of $\Delta{\tau}=100$, during which particles are aggregated into a large filament in the case without obstacles~$N_o=0$, or into a dense nematic loop in the case with obstacles~$N_o=30$. %
Results are shown in~\ref{fig:6}a. %
Without obstacles, $P(m)$ shows as a delta distribution around~$N=900$, as well as an exponential distribution for clusters of small size~$m<10$. %
With obstacles, a delta distribution is shown around~$N_o=600$. %
We also obtain the marginal displacement probability density~$\mathcal{P}(x)$, where each displacement is measured during a time interval of~$\Delta{t}=15$. %
Results are shown in~\ref{fig:6}b. %
Without obstacles~$N_o=0$,~$\mathcal{P}$ shows to be a uniform distribution. %
With obstacles~$N_o=30$, the marginal displacement PDF shows a quadratic dependence~$\mathcal{P}\sim{x^2}$ for small displacements which correspond to the trapped particles in the nematic loop. %
Moreover, for larger displacements a Gaussian-like dependence~$P\sim\exp(-x^2)$ is shown. %

\subsection{Nematic bands}

\begin{figure}[t!]
    \includegraphics[width=\linewidth]{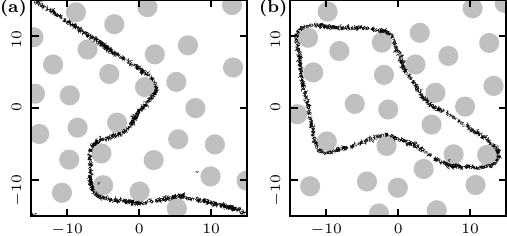}
    \caption{\label{fig:7}%
        Representative snapshots of nematic bands formed for two different obstacle configurations where the band is %
        (a)~percolated, and %
        (b)~forming a closed loop around several obstacles. %
    }
\end{figure}

In a different window of parameter space, active particles instead of self-organizing into polar filaments, they do it into \emph{nematic bands}  ~\cite{barberis2016largescale}. %
To observe nematic bands, the vision cone has to be broader, e.g. $\alpha=0.6\pi$, and rotational diffusion should be enhanced, $\sqrt{D_\theta}=0.8$ with $v_0=1$ ($\mathrm{P_e}\approx{2}$). %
Over time, nematic bands tend to rectify and percolate, as described in Ref.~\cite{barberis2016largescale}. %
In the presence of obstacles, a nematic band is able to avoid obstacles and finds its way until it percolates, see movie in~\cite{SM}. %
Eventually, a nematic  band will close onto itself, forming a nematic loop enclosing multiple obstacles, see~Fig.~\ref{fig:7}. %

\begin{figure}[t!]
    \centering
    \includegraphics[width=\linewidth]{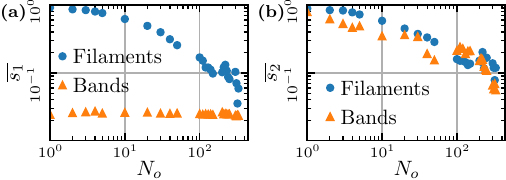}
    \caption{\label{fig:8}%
        Steady state local order parameters for bands and filaments with varying number of obstacles~$N_o$. %
        (a)~Local polar order. %
        (b)~Local nematic order. %
    }
\end{figure}

To further characterize the orientational order of closed loops, we obtain time averages of $s_1$ and~$s_2$ given by~Eq.~\eqref{eq:polar_nematic_order_parameter}. %
For parameters~($\alpha,\sqrt{2D_\theta}$)=($0.3\pi$,0.3) corresponding to filaments, we observe that~$\overline{s_1}$ monotonously decreases with the number of obstacles~$N_o$, see~Fig.~\ref{fig:8}a. %
Without obstacles, particles aggregate into a single dense filament, and with increasing number of obstacles the filament has a larger probability of fragmenting, which explains the decrease in polar order. %
For ($\alpha,\sqrt{2D_{\theta}}$)=($0.6\pi$,0.8) corresponding to bands, $\overline{s_1}$ remains zero at any~$N_o$, as in this case the bands are nematic due to the local interaction between particles. %
Furthermore, the local nematic order parameter~$\overline{s_2}$ shows a monotonic decrease for both, filaments and bands, see~Fig.~\ref{fig:8}b. %
Note that local polar order contributes to the overall value of~$\overline{s_2}$. %
For this reason, filaments show the same behavior for both polar and nematic order. %
For bands, $\overline{s_2}$ is non-vanishing and shows to decrease with increasing~$N_o$, indicating the effect of obstacles is to diminish local nematic order due to the bending around obstacles. %


\section{Summary and conclusions}

Active particles interacting via a vision cone, specifically by non-reciprocal attraction, can form either polar or nematic filaments depending on the vision cone angle~($\alpha$) and the P\'eclet number ($\mathrm{P_e}$)~\cite{barberis2016largescale}. %
Both, polar and nematic filaments grow over time by collecting new active particles. %
Eventually these structure can form closed loops. %
Closed loops, as they form, start shrinking -- not by losing particles but by decreasing the (local) radius of curvature -- until they break up and the structure disintegrates. %
For this reason, closed loops are, in homogeneous environments, short-lived structures that do not play a major role in the large-scale properties of emergent patterns. %
However, here we have shown that, in sharp contrast, in heterogeneous environments, closed loops dominate the large-scale properties of these active systems. %
Closed loops of both, polar and nematic filaments enclosing obstacles, shrink until they enter in contact with such obstacles. %
At this point, these structures stop shrinking and the shape of the loop is stabilized. %
Importantly, these long-lived structures are able to absorb a large fraction of active particles, acting as effective aggregation centers. %
Furthermore, we find that initially polar closed loops become nematic over time as new particles  join the loop, and also due to fluctuations of the orientation vector. %
This implies that, asymptotically, all closed loops become nematic. %

Our study is the first investigation of active particles with non-reciprocal interactions in heterogeneous environments, and reveals that the here-described self-trapping mechanism dominates the large-scale properties of such systems. %
Furthermore, the reported mechanism is likely to be present in many real-world active systems in heterogeneous environments, including animal systems interacting through visual perception, as well as of colloidal and bacterial systems displaying non-reciprocal interactions. %

\begin{acknowledgements}
    R.S was financially supported by the CONACYT-DAAD scholarship program. %
    F.P. acknowledges financial support from C.Y. Initiative of Excellence (grant Investissements d’Avenir ANR-16-IDEX- 0008), INEX 2021 Ambition Project CollInt, Labex MME-DII, projects 2021-258 and 2021-297, and ANR-22-CE30 grant "Push-pull". %
\end{acknowledgements}

\end{document}